\documentclass[11pt]{article}
\usepackage{latexsym}
\usepackage{amssymb}
\usepackage{graphics,graphicx}
\usepackage{tabularx}
\usepackage{amsfonts}
\usepackage{amsmath}
\usepackage{enumerate}
\usepackage{multirow}

\usepackage[dvips]{color}
\usepackage{amsmath}
\usepackage{booktabs}
\usepackage{amsfonts}
\usepackage{multirow}
\usepackage{setspace}
\newcommand{\footmsg}[1]{%
  \let\temp\thempfn%
  \def\thempfs{}
  \footnotetext{#1}
  \let\tempfn\temp}

\begin{document}

\def\ds{\displaystyle}

\newcommand{\beq}{\begin{equation}}
\newcommand{\eeq}{\end{equation}}
\newcommand{\lb}{\label}
\newcommand{\beqar}{\begin{eqnarray}}
\newcommand{\eeqar}{\end{eqnarray}}
\newcommand{\barr}{\begin{array}}
\newcommand{\earr}{\end{array}}
\newcommand{\jump}{\parallel}

\def\c{{\circ}}

\newcommand{\Ehat}{\hat{E}}
\newcommand{\That}{\hat{\bf T}}
\newcommand{\Ahat}{\hat{A}}
\newcommand{\chat}{\hat{c}}
\newcommand{\shat}{\hat{s}}
\newcommand{\khat}{\hat{k}}
\newcommand{\muhat}{\hat{\mu}}
\newcommand{\mc}{M^{\scriptscriptstyle C}}
\newcommand{\mei}{M^{\scriptscriptstyle M,EI}}
\newcommand{\mec}{M^{\scriptscriptstyle M,EC}}
\newcommand{\hbeta}{{\hat{\beta}}}
\newcommand{\rec}[2]{\left( #1 #2 \ds{\frac{1}{#1}}\right)}
\newcommand{\rep}[2]{\left( {#1}^2 #2 \ds{\frac{1}{{#1}^2}}\right)}
\newcommand{\derp}[2]{\ds{\frac {\partial #1}{\partial #2}}}
\newcommand{\derpn}[3]{\ds{\frac {\partial^{#3}#1}{\partial #2^{#3}}}}
\newcommand{\dert}[2]{\ds{\frac {d #1}{d #2}}}
\newcommand{\dertn}[3]{\ds{\frac {d^{#3} #1}{d #2^{#3}}}}

\def\bob{{\, \underline{\overline{\otimes}} \,}}
\def\ob{{\, \underline{\otimes} \,}}
\def\scalp{\mbox{\boldmath$\, \cdot \, $}}
\def\gdp{\makebox{\raisebox{-.215ex}{$\Box$}\hspace{-.778em}$\times$}}
\def\daa{\makebox{\raisebox{-.050ex}{$-$}\hspace{-.550em}$: ~$}}
\def\mK{\mbox{${\mathcal{K}}$}}
\def\cK{\mbox{${\mathbb {K}}$}}

\def\Xint#1{\mathchoice
   {\XXint\displaystyle\textstyle{#1}}%
   {\XXint\textstyle\scriptstyle{#1}}%
   {\XXint\scriptstyle\scriptscriptstyle{#1}}%
   {\XXint\scriptscriptstyle\scriptscriptstyle{#1}}%
   \!\int}
\def\XXint#1#2#3{{\setbox0=\hbox{$#1{#2#3}{\int}$}
     \vcenter{\hbox{$#2#3$}}\kern-.5\wd0}}
\def\ddashint{\Xint=}
\def\fpint{\Xint=}
\def\dashint{\Xint-}
\def\cpvint{\Xint-}
\def\intl{\int\limits}
\def\cpvintl{\cpvint\limits}
\def\fpintl{\fpint\limits}
\def\ointl{\oint\limits}
\def\bA{{\bf A}}
\def\ba{{\bf a}}
\def\bB{{\bf B}}
\def\bb{{\bf b}}
\def\bc{{\bf c}}
\def\bC{{\bf C}}
\def\bD{{\bf D}}
\def\bE{{\bf E}}
\def\be{{\bf e}}
\def\bbf{{\bf f}}
\def\bF{{\bf F}}
\def\bG{{\bf G}}
\def\bg{{\bf g}}
\def\bi{{\bf i}}
\def\bH{{\bf H}}
\def\bK{{\bf K}}
\def\bL{{\bf L}}
\def\bM{{\bf M}}
\def\bN{{\bf N}}
\def\bn{{\bf n}}
\def\bm{{\bf m}}
\def\b0{{\bf 0}}
\def\bo{{\bf o}}
\def\bX{{\bf X}}
\def\bx{{\bf x}}
\def\bP{{\bf P}}
\def\bp{{\bf p}}
\def\bQ{{\bf Q}}
\def\bq{{\bf q}}
\def\bR{{\bf R}}
\def\bS{{\bf S}}
\def\bs{{\bf s}}
\def\bT{{\bf T}}
\def\bt{{\bf t}}
\def\bU{{\bf U}}
\def\bu{{\bf u}}
\def\bv{{\bf v}}
\def\bw{{\bf w}}
\def\bW{{\bf W}}
\def\by{{\bf y}}
\def\bz{{\bf z}}
\def\T{{\bf T}}
\def\Te{\textrm{T}}
\def\Id{{\bf I}}
\def\bxi{\mbox{\boldmath${\xi}$}}
\def\balpha{\mbox{\boldmath${\alpha}$}}
\def\bbeta{\mbox{\boldmath${\beta}$}}
\def\bepsilon{\mbox{\boldmath${\epsilon}$}}
\def\bvarepsilon{\mbox{\boldmath${\varepsilon}$}}
\def\bomega{\mbox{\boldmath${\omega}$}}
\def\bphi{\mbox{\boldmath${\phi}$}}
\def\bsigma{\mbox{\boldmath${\sigma}$}}
\def\bfeta{\mbox{\boldmath${\eta}$}}
\def\bDelta{\mbox{\boldmath${\Delta}$}}
\def\btau{\mbox{\boldmath $\tau$}}
\def\tr{{\rm tr}}
\def\dev{{\rm dev}}
\def\div{{\rm div}}
\def\Div{{\rm Div}}
\def\Grad{{\rm Grad}}
\def\grad{{\rm grad}}
\def\Lin{{\rm Lin}}
\def\Sym{{\rm Sym}}
\def\Skw{{\rm Skew}}
\def\abs{{\rm abs}}
\def\Re{{\rm Re}}
\def\Im{{\rm Im}}
\def\capB{\mbox{\boldmath${\mathsf B}$}}
\def\capC{\mbox{\boldmath${\mathsf C}$}}
\def\capD{\mbox{\boldmath${\mathsf D}$}}
\def\capE{\mbox{\boldmath${\mathsf E}$}}
\def\capG{\mbox{\boldmath${\mathsf G}$}}
\def\tcapG{\tilde{\capG}}
\def\capH{\mbox{\boldmath${\mathsf H}$}}
\def\capK{\mbox{\boldmath${\mathsf K}$}}
\def\capL{\mbox{\boldmath${\mathsf L}$}}
\def\capM{\mbox{\boldmath${\mathsf M}$}}
\def\capR{\mbox{\boldmath${\mathsf R}$}}
\def\capW{\mbox{\boldmath${\mathsf W}$}}

\def\i{\mbox{${\mathrm i}$}}
\def\mC{\mbox{\boldmath${\mathcal C}$}}
\def\mB{\mbox{${\mathcal B}$}}
\def\mE{\mbox{${\mathcal{E}}$}}
\def\mL{\mbox{${\mathcal{L}}$}}
\def\mK{\mbox{${\mathcal{K}}$}}
\def\mV{\mbox{${\mathcal{V}}$}}
\def\C{\mbox{\boldmath${\mathcal C}$}}
\def\E{\mbox{\boldmath${\mathcal E}$}}

\def\ACME{{ Arch. Comput. Meth. Engng.\ }}
\def\ARMA{{ Arch. Rat. Mech. Analysis\ }}
\def\AMR{{ Appl. Mech. Rev.\ }}
\def\ASCEEM{{ ASCE J. Eng. Mech.\ }}
\def\acta{{ Acta Mater. \ }}
\def\CMAME {{ Comput. Meth. Appl. Mech. Engrg.\ }}
\def\CRAS{{ C. R. Acad. Sci., Paris\ }}
\def\EFM{{ Eng. Fract. Mech.\ }}
\def\EJMA{{ Eur.~J.~Mechanics-A/Solids\ }}
\def\IJES{{ Int. J. Eng. Sci.\ }}
\def\IJF{{\it Int. J. Fracture}}
\def\IJMS{{ Int. J. Mech. Sci.\ }}
\def\IJNAMG{{ Int. J. Numer. Anal. Meth. Geomech.\ }}
\def\IJP{{ Int. J. Plasticity\ }}
\def\IJSS{{ Int. J. Solids Structures\ }}
\def\IngA{{ Ing. Archiv\ }}
\def\JAM{{ J. Appl. Mech.\ }}
\def\JAP{{ J. Appl. Phys.\ }}
\def\JE{{ J. Elasticity\ }}
\def\JM{{ J. de M\'ecanique\ }}
\def\JMPS{{ J. Mech. Phys. Solids\ }}
\def\Macro{{ Macromolecules\ }}
\def\MOM{{ Mech. Materials\ }}
\def\MMS{{ Math. Mech. Solids\ }}
\def\MMT{{\it Metall. Mater. Trans. A}}
\def\MPCPS{{ Math. Proc. Camb. Phil. Soc.\ }}
\def\MSE{{ Mater. Sci. Eng.}}
\def\PMPS{{ Proc. Math. Phys. Soc.\ }}
\def\PRE{{ Phys. Rev. E\ }}
\def\PRSL{{ Proc. R. Soc.\ }}
\def\rock{{ Rock Mech. and Rock Eng.\ }}
\def\QAM{{ Quart. Appl. Math.\ }}
\def\QJMAM{{ Quart. J. Mech. Appl. Math.\ }}
\def\SCRMAT{{ Scripta Mater.\ }}
\def\SM{{\it Scripta Metall. }}


\def\salto#1#2{
[\mbox{\hspace{-#1em}}[#2]\mbox{\hspace{-#1em}}]}




\title{
Stress concentration near stiff inclusions:\\
validation of rigid inclusion model and boundary layers \\ by means of photoelasticity}

\author{D. Misseroni, F. Dal Corso,  S. Shahzad, D. Bigoni$^0$ \\
University of Trento, via Mesiano 77, I-38123 Trento, Italy \\
}
\date{}
\maketitle
\footnotetext[0]{
Corresponding author:~Davide~Bigoni - fax:~+39~0461~882599; tel.:~+39~0461~882507; web-site:~http://www.ing.unitn.it/$\sim$bigoni/;
e-mail:~bigoni@ing.unitn.it. Additional e-mail addresses: diego.misseroni@ing.unitn.it (Diego Misseroni),
francesco.dalcorso@ing.unitn.it (Francesco Dal Corso), summer.shahzad@unitn.it (Summer Shahzad).}

\begin{abstract}
Photoelasticity is employed to investigate the stress state near stiff rectangular and  rhombohedral
inclusions embedded in a \lq soft' elastic plate.
Results show that the singular stress field predicted by the linear elastic solution for the rigid inclusion model
can be generated in reality, with great accuracy, within a material.
In particular, experiments:
(i.) agree with the fact that the  singularity is lower for obtuse than for acute inclusion angles;  (ii.) show that
the singularity is stronger in Mode II than in Mode I (differently from a notch);
(iii.) validate the model of rigid quadrilateral inclusion;
(iv.) for thin inclusions, show the presence of boundary layers
deeply influencing the stress field, so that the limit case of rigid line inclusion is obtained
in strong dependence on the inclusion's shape.
The introduced experimental methodology opens the possibility of
enhancing the design of thin reinforcements and of
analyzing
complex situations involving interaction between inclusions and defects.

\end{abstract}


\section{Introduction}

Experimental stress analysis near a crack or a void has been the subject of an intense research effort (see for instance Lim and Ravi-Chandar, 2007; 2009; Schubnel et al. 2011; Templeton et al. 2009), but the stress field near a
rigid inclusion embedded in an elastic matrix, a fundamental problem in the design of composites,
has surprisingly been left almost unexplored
(Theocaris, 1975; Theocaris and Paipetis, 1976 a; b; Reedy and Guess, 2001) and has {\it never} been investigated via
photoelasticity\footnote{Gdoutos (1982) reports plots of the fields that would
result from photoelastic investigation of rigid cusp inclusions, but does not report any experiment,while
Theocaris and Paipetis (1976b) show only one photo of very low quality for a rigid line inclusion.
Noselli et al. (2011) (see also Bigoni, 2012; Dal Corso et al. 2008) only treat the case of a thin line-inclusion.
Theocaris and Paipetis (1976 a;b) use the method of caustics (see also Rosakis and Zehnder, 1985). This method, suited for
determining the stress intensity factor, suffers from the drawback that near the boundary of a stiff inclusion the state of strain can be
closer to plane strain than to plane stress, a feature affecting the shape of the caustics.
}.

Though the analytical determination of elastic fields
around inclusions is a problem in principle solvable with existing
methodologies (Movchan and Movchan, 1995; Muskhelishvili, 1953; Savin, 1961),
detailed treatments
are not available and the existing solutions\footnote{
Evan-Iwanowski (1956) treated the case of a triangular rigid inclusion, Chang and Conway (1968) addressed rectangular rigid inclusions, while
Panasyuk et al. (1972) considered the problem of
the stress distribution in the neighborhood of a cuspidal point of a rigid inclusion
embedded in a matrix. Ishikawa and Kohno (1993) and Kohno and Ishikawa (1994)
developed a method for the calculation of the stress singularity orders and the
stress intensities at a singular point in an polygonal inclusion.
} lack mechanical interpretation,
in the sense that it is not known if these predict stress fields observable in reality\footnote{
The experimental methodology introduced in the present article for rigid inclusions can be of interest for the
experimental investigation of the interaction between inclusions and defects, such as for instance cracks or shear bands, for which
analytical solutions are already available (for cracks, see Piccolroaz et al. 2012 a; b; Valentini et al. 1999, while for shear bands, see
Dal Corso and Bigoni 2009, Dal Corso and Bigoni 2010).
}. Moreover, from experimental point of view, questions arise whether the bonding between inclusion and matrix can be realized
and can resist loading without detachment (which would introduce a crack) and if
self-stresses can be reduced to negligible values.
In this article we (i.) re-derive asymptotic and full-field solutions
for rectangular and  rhombohedral rigid inclusions (Section \ref{teoria}) and (ii.) compare
these with photoelastic experiments (Section \ref{pratica}).
\begin{figure}[!htcb]
  \begin{center}
\includegraphics[width=13 cm]{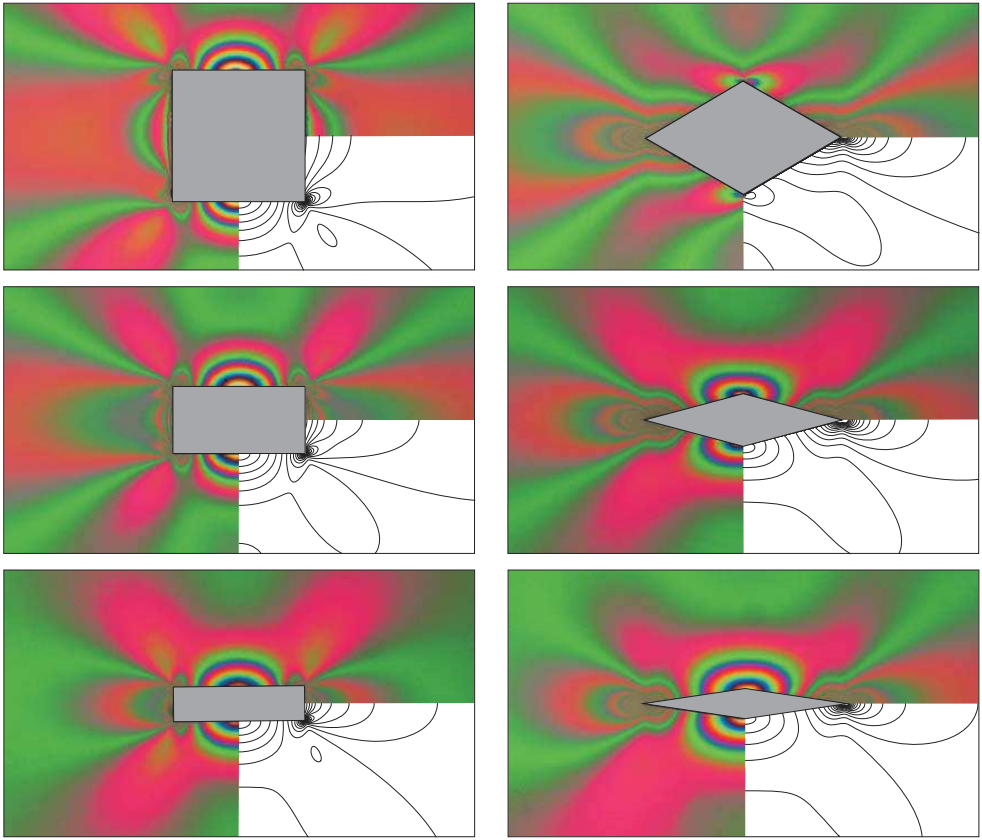}
\caption{
Photoelastic fringes revealing the stress field near stiff
(made up of polycarbonate, Young modulus 2350 MPa)
rectangular (large edge $l_x=$20 mm, edges aspect ratios $l_y/l_x=\{1;1/2;1/4\}$) and rhombohedral (large axis $l_x=$30 mm,
axis aspect ratios $l_y/l_x=\{9/15;4/15;2/15\}$)  inclusions
embedded in an elastic
matrix (a two-component epoxy resin, Young modulus 22 MPa, approximatively 100 times less stiff than the inclusions)
and loaded with a remote uniaxial tensile stress $\sigma_{xx}^\infty=$0.28 MPa, compared to the elastic solution for rigid inclusions (in plane stress, with Poisson's ratio equal to 0.49).
}
\label{strisciata}
 \end{center}
\end{figure}

Photoelastic fringes obtained with a white circular polariscope are shown in Fig. \ref{strisciata} and
indicate that the linear elastic solutions provide an excellent description of the elastic fields
generated by inclusions up to a distance so close to the edges of the inclusions that fringes result
unreadable (even with the aid of an optical microscope). By comparison of the photos shown in Fig. \ref{strisciata} with Fig. 1
of Noselli et al. (2010), it can be noted that the stress fields correctly tend to those relative to a rigid line inclusion ({\it stiffener}),
when the aspect ratio of the inclusions decreases, and
that the stress fields very close to a thin inhomogeneity are substantially affected by boundary layers depending on
the (rectangular or rhombohedral) shape.

\section{Theoretical linear elastic fields near rigid polygonal inclusions} \lb{teoria}

The stress/strain fields in a linear isotropic elastic matrix containing a rigid polygonal inclusion are obtained
analytically through both an asymptotic approach and a full-field determination.
Considering plane stress or strain conditions,  the displacement components in the $x-y$ plane are
\beq\label{displ}
u_x=u_x(x,y), \qquad u_y=u_y(x,y),
\eeq
corresponding to the following in-plane deformations $\varepsilon_{\alpha\beta}$ ($\alpha$, $\beta$=$x$, $y$)
\beq
\label{defcomp}
\varepsilon_{xx}=u_{x,x}, \qquad
\varepsilon_{yy}=u_{y,y}, \qquad
\varepsilon_{xy}=\frac{u_{x,y}+u_{y,x}}{2},
\eeq
which, for linear elastic isotropic behaviour, are related to the in the in-plane stress components $\sigma_{\alpha\beta}$
 ($\alpha$, $\beta$=$x$, $y$) via
\beq
\lb{coscosti}
\ds\varepsilon_{xx}=\frac{(\kappa+1)\sigma_{xx}+(\kappa-3)\sigma_{yy}}{8\mu }, \qquad
\ds\varepsilon_{yy}=\frac{(\kappa+1)\sigma_{yy}+(\kappa-3)\sigma_{xx}}{8\mu }, \qquad
\ds\varepsilon_{xy}=\frac{\sigma_{xy}}{2\mu },
\eeq
where $\mu$ represents the shear modulus and
$\kappa\geq 1$ is equal to $3-4 \nu$ for plane strain or $(3-\nu)/(1+\nu)$ for plane stress,
where $\nu\in(-1,1/2)$ is the Poisson's ratio.
Finally, in the absence of body forces, the in-plane stresses satisfy the equilibrium equation (where repeated indices are summed)
\beq
\lb{equilibrio}
\sigma_{\alpha\beta, \beta}= 0.
\eeq

\subsection{Asymptotic fields near the corner of a rigid wedge}\label{asintotototo}

\begin{figure}[!htcb]
  \begin{center}
\includegraphics[width=13 cm]{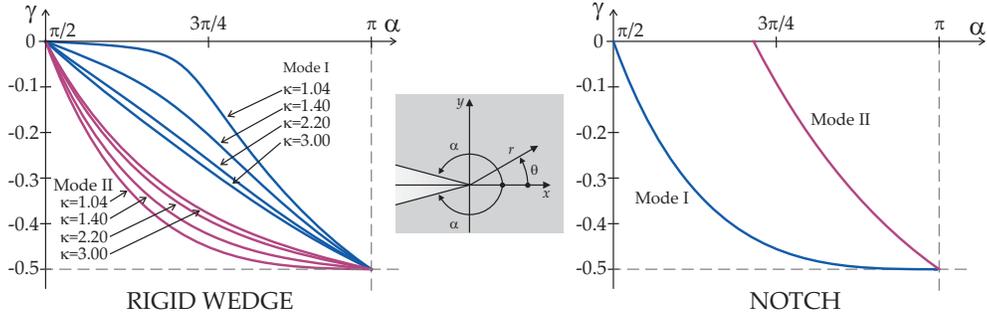}
\caption{The higher singularity power $\gamma$ for a rigid wedge
(left, angle $\alpha$ is the semi-angle in the matrix enclosing the wedge) and for a notch (right, angle $\alpha$ is the semi-angle in the matrix enclosing the notch) under Mode I and Mode II loading and different values of $\kappa$. }
\label{singolarita}
 \end{center}
\end{figure}

Near the corner of a rigid wedge the mechanical fields may be
approximated by their asymptotic expansions (Williams, 1952). With reference to the polar coordinates $r,\vartheta$ centered
at the wedge corner and such that the elastic matrix occupies the region $\vartheta\in[-\alpha, \alpha]$
(while the semi-infinite rigid wedge lies in the remaining part of plane, Fig. \ref{singolarita}),
the Airy function $F(r,\vartheta)$, automatically satisfying the equilibrium equation (\ref{equilibrio}), is defined as
\beq
\sigma_{rr}=\ds\frac{1}{r}\left(F_{,r}+\frac{F_{,\vartheta \vartheta}}{r}\right), \qquad
\sigma_{\vartheta\vartheta}=\ds F_{,rr},\qquad
\sigma_{r\vartheta}=-\ds\left(\frac{ F_{,\vartheta}}{r}\right)_{,r}.
\eeq
The following power-law form of the Airy function satisfies the kinematic compatibility conditions [Barber, 1993, his eq. (11.35)]
\beq
F(r,\vartheta)= r^{\gamma+2}
\left[A_1\cos(\gamma+2)\vartheta+A_2 \sin(\gamma+2)\vartheta+A_3\cos\gamma\vartheta +A_4 \sin\gamma\vartheta\right],
\eeq
and provides the in-plane stress components as
 \beq\barr{cll}
\sigma_{rr}&=&
-(\gamma+1)r^{\gamma}[A_1(\gamma+2)\cos(\gamma+2)\vartheta+A_2(\gamma+2)\sin(\gamma+2)\vartheta\\[4mm]
&&+A_3(\gamma-2)\cos\gamma\vartheta
+A_4(\gamma-2)\sin\gamma\vartheta],\\ [6mm]
\sigma_{\vartheta\vartheta}&=&
(\gamma+2)(\gamma+1)r^{\gamma}[A_1\cos(\gamma+2)\vartheta+A_2\sin(\gamma+2)\vartheta\\[4mm]
&&+A_3\cos\gamma\vartheta+A_4\sin\gamma\vartheta],\\[6mm]
\sigma_{r\vartheta}&=&
(\gamma+1)r^{\gamma}[A_1(\gamma+2)\sin(\gamma+2)\vartheta-A_2(\gamma+2)\cos(\gamma+2)\vartheta\\[4mm]
&&+A_3\gamma\sin\gamma\vartheta-A_4\gamma\cos\gamma\vartheta],
\earr
\eeq
where $A_1,A_2$ and $A_3,A_4$ are unknown constants defining
the symmetric (Mode I) and antisymmetric (Mode II) contributions, respectively,
while $\gamma$ represents the unknown power of $r$ for the stress and strain asymptotic fields,
$\left\{\sigma_{\alpha\beta},\varepsilon_{\alpha\beta}\right\}\sim r^\gamma$,
with $\gamma\geq-1/2$.

Imposing the boundary displacement conditions
$u_r(r,\pm\alpha)= u_\vartheta(r,\pm\alpha)=0$
leads to two decoupled homogeneous systems,
one for each Mode symmetry condition, so that non-trivial asymptotic fields
are obtained when determinant of coefficient matrix is null, namely (Seweryn and Molski, 1996)
\beq\label{detnullo}
\barr{lll}
&(\gamma+1)\sin(2\alpha)-\kappa\sin(2\alpha(\gamma+1))=0,\qquad &\mbox{Mode I};\\[4mm]
&(\gamma+1)\sin(2\alpha)+\kappa\sin(2\alpha(\gamma+1))=0, \qquad &\mbox{Mode II}.
\earr
\eeq
Note that, in the limit $\kappa=1$ (incompressible material under plane strain conditions),
equations (\ref{detnullo}) are the same as those obtained for a notch, except that the loading Modes are
switched. Furthermore, according to the so-called \lq Dundurs correspondence' (Dundurs, 1989), when $\kappa=-1$ eqns
(\ref{detnullo}) coincide with those corresponding to a notch.

The smallest negative value of the power $\gamma\geq-1/2$ for each loading Mode, satisfying eqn (\ref{detnullo})$_1$ and (\ref{detnullo})$_2$,
represents the leading order term
of the asymptotic expansion.
These two values (one for Mode I and another for Mode II) are reported in
Fig. \ref{singolarita} (left), for different values of $\kappa$, as functions of the semi-angle $\alpha$ and compared with the respective values
for a void wedge, Fig. \ref{singolarita} (right).

For the rigid wedge, \emph{similarly to the notch problem}:
\begin{itemize}
\item  the singularity appears only when $\alpha > \pi/2$ and  increases with the increase of $\alpha$;
\item  a square root singularity ($\sigma_{\alpha\beta}\sim 1/\sqrt{r}$) appears for both mode I and II
when $\alpha$ approaches $\pi$ (corresponding to the rigid line inclusion model, see Noselli et al. 2010);
\end{itemize}
while, \emph{differently from the notch problem}:
\begin{itemize}
\item the singularity depends on the Poisson's ratio $\nu$ through the parameter $\kappa$;
\item the singularity under Mode II condition is stronger than that under Mode I; in particular,
a weak singularity is developed under Mode I
when, for plane strain deformation, a quasi-incompressible material ($\nu$ close to $1/2$) contains a
rigid wedge with $\alpha \in [\frac{1}{2},\frac{3}{4}]\pi$.
\end{itemize}

Since the intensity of singularity near a corner
is strongly affected by the value of the angle $\alpha$, it follows that
the stress field close to a rectangular inclusion is substantially different to that close to a rhombohedral one.
Therefore, strongly different boundary layers arise when a rectangular or a rhombohedral inclusion approaches the limit of line inclusion.

\subsection{Full-field solution for a matrix containing a polygonal rigid inclusion}\label{campopieno}

Solutions in 2D isotropic elasticity can be obtained using the method of complex potentials (Muskhelishvili, 1953),
where the generic point ($x,y$) is referred to the complex variable $z=x+\i\,y$ (where $\i$ is the imaginary unit) and
the mechanical fields are given in terms of complex potentials $\varphi(z)$ and $\psi(z)$ which can be computed from the
boundary conditions.

In the case of non-circular inclusions, it is instrumental
to introduce the complex variable $\zeta$, related to the physical plane
through $ z=\omega(\zeta)$ with  the conformal mapping function $\omega$ (such that the inclusion boundary becomes
a unit circle in the $\zeta$-plane, $\zeta=e^{\i \theta}$), so that
 the stress and displacement components are given as
\beq
\left\{\barr{l}
\ds \sigma_{xx}+\sigma_{yy}=
4\Re\left[\frac{\varphi'(\zeta)}{\omega'(\zeta)}\right],\\[6mm]
\ds \sigma_{yy}-\sigma_{xx}+2\,\i\, \sigma_{xy}=2\left[\frac{\psi'(\zeta)}{\omega'(\zeta)}+\frac{\overline{\omega(\zeta)}}{\omega'(\zeta)^3}
\left[\varphi''(\zeta)\omega'(\zeta)-\varphi'(\zeta)\omega''(\zeta)\right]\right],\\[6mm]
\ds 2\mu(u_x+\,\i\, u_y)=\kappa\varphi(\zeta)-\frac{\omega(\zeta)}{\overline{\omega'(\zeta)}}\overline{\varphi'(\zeta)}-\overline{\psi(\zeta)}.
\earr
\right.
\eeq
The complex potentials are the sum of the unperturbed (homogeneous) solution
and the perturbed (introduced by the inclusion) solution, so that, considering
boundary conditions at infinity of constant stress with the only non-null component $\sigma_{xx}^\infty$,
we may write
\beq
\ds\varphi(\zeta)=\frac{\sigma_{xx}^\infty}{4}\omega(\zeta)+\varphi^{(p)}(\zeta), \qquad
\ds\psi(\zeta)=-\frac{\sigma_{xx}^\infty}{2}\omega(\zeta)+\psi^{(p)}(\zeta),
\eeq
where the perturbed potentials $\varphi^{(p)}(\zeta)$ and $\psi^{(p)}(\zeta)$ can be
obtained by imposing the conditions on the inclusion boundary, which are defined
on a unit circle and  for a rigid inclusion\footnote{
Eqn (\ref{bcrigid}) holds when rigid-body displacements are excluded.
} are
\beq\label{bcrigid}
\ds\kappa\varphi^{(p)}(\zeta)-\frac{\omega(\zeta)}{\overline{\omega'(\zeta)}}\overline{{\varphi^{(p)}}'(\zeta)}
-\overline{\psi^{(p)}(\zeta)}=\frac{\sigma_{xx}^\infty}{2}\left(\frac{1-\kappa}{2}\omega(\zeta)-\overline{\omega(\zeta)}\right),
\qquad \mbox{for}\,\,\zeta=e^{\i \theta},
\,\,\theta\in [0,2\pi].
\eeq
In the case of $n$-polygonal shape inclusions the conformal mapping which maps
the interior of the unit disk onto the region exterior to the inclusion
is given by the Schwarz-Christoffel integral
\beq\label{SC}
\omega(\zeta)=\ds R e^{i\alpha_0}\int^{\zeta}_{1}\prod^{n}_{j=1}\left(1-\frac{s}{k_j}\right)^{1-\alpha_j}\frac{ds}{s^2}+k_0,
\eeq
 where $R$, $k_0$, and $\alpha_0$ are constants representing scaling, translation, and rotation of the inclusion,
while $k_j$ and $\alpha_j$ ($j$=1,..., $n$)
are the pre-images of the $j$-th vertex in the $\zeta$ plane and the fraction of $\pi$ of the  $j$-th interior angle, respectively.
In the following the translation and rotation parameters for the inclusion are taken null, $k_0=\alpha_0=0$.

Assuming that the perturbed potentials are holomorphic
inside the unit circle in the $\zeta$-plane, $\varphi^{(p)}(\zeta)$ can be
expressed through Laurent series
\beq
 \ds\varphi^{(p)}(\zeta)=R\,\sigma_{xx}^\infty \sum^{\infty}_{j=1}a_j\zeta^j,
\eeq
where $a_j$ ($j$=1,2,3,...) are unknown complex constants.
Furthermore, since the integral expression in eqn (\ref{SC}) cannot be computed  as closed form for generic polygon, it is expedient to represent
the conformal mapping as
\beq
\omega(\zeta)=R\left(\frac{1}{\zeta}+\sum^{\infty}_{j=1} d_j\zeta^j\right),
\eeq
where $d_j$ ($j$=1,2,3,...) are complex constants.

In order to obtain an approximation for the solution, the series expansions for $\omega(\zeta)$ and $\varphi^{(p)}(\zeta)$
are truncated at the $M$-th term. Through Cauchy integral theorem,
integration over the inclusion boundary of eqn (\ref{bcrigid})
yields a linear system for the $M$ unknown complex constants $a_j$,
functions of the $M$ constants $d_j$,  obtained through series expansion of eqn. (\ref{SC}). Once the expression
for $\varphi^{(p)}(\zeta)$
is obtained, the integral over the inclusion boundary of the conjugate version of the boundary condition (\ref{bcrigid}) is used to approximate
$\psi^{(p)}(\zeta)$, resulting as
\beq
\psi^{(p)}(\zeta) = \frac{\ds\sum_{j=1}^{M+2} b_j\zeta^{j-1}}{\ds\sum_{j=1}^{M+2} c_j\zeta^{j-1}}\,\,\ds R\,\sigma_{xx}^\infty\, \zeta.
\eeq

\paragraph{Rectangle}

In this case the angle fractions are $\alpha_j=1/2$ ($j$=1,..., $4$) while the pre-images are
\beq
\ds k_1=e^{\eta \pi \i}, \qquad
\ds k_2=e^{-\eta  \pi \i}, \qquad
\ds k_3=e^{(1+\eta)\pi \i}, \qquad
\ds k_4=e^{(1-\eta)\pi \i},
\eeq
where $\eta$ (likewise $R$) is a parameter function of the rectangle aspect ratio $l_y/l_x$, with the inclusion edges $l_x$ and $l_y$.
Parameters $\eta$ and $R$ are given in Tab. 1
for the aspect ratios considered here.

\begin{table} [!htcb]
\begin{center} \begin{tabular}{c|ccc}
\toprule[.8pt]
$l_y/l_x$&\bf{1}&\bf{1/2}&\bf{1/4}\\
\hline
$\eta$&  0.2500 & 0.2003 & 0.1548\\
$R/l_x$& 0.5902 & 0.4374 & 0.3539\\
\bottomrule
\end{tabular}
\label{rettangolovalori}
\bf \small \caption{\textnormal{Parameters $\eta$ and $R$ for the considered aspect ratios $l_y/l_x$ of rectangular rigid inclusions.
}}
\end{center}
\end{table}

The conformal mapping function and perturbed potentials obtained in the case of rectangle with $l_y/l_x=1/4$
are reported for $M$=15:
\beq\begin{array}{rll}
\omega(\zeta)=&\left(\ds\frac{1}{\zeta}+
0.5633\zeta-0.1138\zeta^3-0.0385\zeta^5-0.0071\zeta^7+0.0042\zeta^9+0.0052\zeta^{11}\right.\\[3mm]
&\left.+0.0022\zeta^{13}-0.0006\zeta^{15}\right)R,\\[5mm]

\varphi^{(p)}(\zeta) =& \left( - 0.2420- 0.0264\zeta^2-
    0.0071\zeta^4+0.0003\zeta^6+0.0020\zeta^8+0.0012\zeta^{10}+0.0002\zeta^{12}\right.\\[3mm]
&\left.- 0.0001\zeta^{14}\right)\,\,\ds R\,\sigma_{xx}^\infty\,\zeta,\\[5mm]

\psi^{(p)}(\zeta) =&
\ds\left(-2.4454-54.9115\zeta^2+6.4081\zeta^4+5.5545\zeta^6+3.4073\zeta^8+0.6051\zeta^{10}-1.3007\zeta^{12}\right.\\[3mm]
&\left.-1.0545\zeta^{14}+0.2727\zeta^{16}\right)
R\,\sigma_{xx}^\infty\, \zeta  / \left(109.8986-61.9012\zeta^2+37.5162\zeta^4+21.1312\zeta^6\right.\\[3mm]
&\left.+5.4989\zeta^8-4.1163\zeta^{10}-6.2272\zeta^{12}-3.1597\zeta^{14}+\zeta^{16}\right).
\earr \eeq

\paragraph{Rhombus}

In this case the pre-images are
\beq
\ds k_1=1, \qquad
\ds k_2=\i, \qquad
\ds k_3=-1, \qquad
\ds k_4=-\i,
\eeq
while the angle fractions are
\beq
\alpha_1=\alpha_3=\frac{2}{\pi}\arctan{(l_y/l_x)},
\qquad
\alpha_2=\alpha_4=1-\alpha_1.
\eeq
The scaling parameter $R$ is reported in Tab. 2
for the rhombus aspect ratios $l_y/l_x$ considered here, where $l_x$ and $l_y$ are the inclusion axis.

\begin{table} [!htcb]\label{rombovalori}
\begin{center} \begin{tabular}{c|ccc}
\toprule[.8pt]
$l_y/l_x$&\bf{9/15}&\bf{4/15}&\bf{2/15}\\
\hline
$R/l_x$& 0.3389 & 0.2841 & 0.2659 \\
\bottomrule
\end{tabular}
\bf \small \caption{\textnormal{Parameter $R$ for the considered aspect ratios $l_y/l_x$ of rhombohedral rigid inclusions.}}
\end{center}
\end{table}

The conformal mapping function and perturbed potentials obtained in the case of rhombus with $l_y/l_x=2/15$
are reported for $M$=15:

\beq\begin{array}{rll}
\omega(\zeta)=&\left(\ds\frac{1}{\zeta}+0.8312\zeta+ 0.0515\zeta^3-0.0086\zeta^5+0.0068\zeta^7- 0.0028\zeta^9+0.0025\zeta^{11}\right.\\[3mm]
&\left.-0.0013\zeta^{13}+0.0013\zeta^{15}\right)R,\\[5mm]

\varphi^{(p)}(\zeta) =& \left(
-0.1628+0.0071\zeta^2+0.0001\zeta^4+0.0009\zeta^6+0.0001\zeta^8+0.0003\zeta^{10}+0.0001\zeta^{12}\right.\\[3mm]
&\left.+0.0002\zeta^{14}\right)\,\,\ds R\,\sigma_{xx}^\infty\,\zeta,\\[5mm]

\psi^{(p)}(\zeta) =& \ds\left(8.1122+
28.1115\zeta^2+1.8150\zeta^4-0.6928\zeta^6+0.4105\zeta^8-0.4451\zeta^{10}+0.1665\zeta^{12}\right.\\[3mm]
&\left.-0.3417\zeta^{14}+0.2727\zeta^{16}\right)
R\,\sigma_{xx}^\infty\, \zeta  / \left(-53.0727+44.1156\zeta^2+8.2012\zeta^4-2.2724\zeta^6\right.\\[3mm]
&\left.+2.5225\zeta^8-1.3283\zeta^{10}
+1.4453\zeta^{12}-0.9307\zeta^{14}+\zeta^{16}\right).
\earr \eeq

\section{Photoelastic elastic fields near rigid polygonal inclusions} \lb{pratica}

Photoelastic experiments with linear and circular polariscope (with quarterwave retarders for 560nm) at white and
monochromatic light\footnote{
The polariscope (dark field arrangement and equipped with a
white and sodium vapor lightbox at $\lambda$  = 589.3nm,
purchased from Tiedemann \& Betz) has been designed by us and manufactured at the University of Trento, see
http://ssmg.unitn.it/
for a detailed description of  the apparatus.
}
have been performed on twelve two-component resin (Translux D180 from Axon; mixing ratio by weight: hardener 95, resin 100,
accelerator 1.5; the elastic modulus of the resulting matrix has been measured by us to be 22 MPa,
while the Poisson's ratio has been indirectly estimated equal to 0.49) samples containing stiff inclusions,
obtained with a solid polycarbonate 3 mm thick sheet (clear 2099 Makrolon UV) from Bayer with elastic modulus equal to 2350 MPa,
approximatively 100 times stiffer than the matrix.

Samples have been prepared by pouring the resin (after deaeration, obtained through a 30 minutes
exposition at a pressure of -1 bar) into a teflon mold (340 mm $\times$ 120 mm $\times$ 10 mm) to obtain 3$\pm$0.05 mm thick samples.
The resin has been kept for 36 hours at constant temperature of 29 $^\circ$C and humidity of 48\%.
After mold extraction, samples have been cut to be 320\,mm $\times$ 110\,mm $\times$ 3\,mm,
containing rectangular inclusions with wedges 20 mm $\times$ $\left\{20; 10; 5\right\}$ mm and
rhombohedral inclusions with axis 30 mm $\times$ $\left\{18; 8; 4\right\}$ mm.

Photos have been taken with a Nikon D200 digital camera,
equipped with a AF-S micro Nikkor (105 mm, 1:2.8G ED) lens and
with a AF-S micro Nikkor (70–180 mm, 1:4.5–5.6 D) lens for details.
Monitoring with a thermocouple
connected to a Xplorer GLX Pasco$^\copyright$, temperature near the samples during experiments has been found  to lie
around 22.5$^\circ$ C, without sensible oscillations.
Near-tip fringes have been captured with a Nikon SMZ800 stereozoom microscope equipped with
Nikon Plan Apo 0.5x objective and a Nikon DS-Fi1 high-definition color camera head.

The uniaxial stress experiments have been performed at controlled vertical load
applied in discrete steps, increasing from 0 to a maximum load of 90 N, except for thin rectangular and rhombohedral inclusions, where the maximum load has been 70 N and 78 N, respectively  (loads have been reduced for thin inclusions to prevent failure at the vertex tips). In all cases an additional load of
3.4N has been applied, corresponding to the grasp weight, so that
maximum nominal far-field stress of 0.28 MPa  has been applied (0.22 MPa and 0.25 MPa for the thin inclusions).

Data have been acquired after 5 minutes from the load application time in order to damp down the largest
amount of viscous deformation, noticed as a settlement of the fringes, which follows displacement stabilization.
Releasing the applied load after the maximum amount, all the samples at rest showed no perceivably residual stresses in the whole specimen.

Comparison between analytical solutions and experiments is possible through matching of the
 isochromatic fringe order  $N$, which (in linear photoelasticity)\footnote{
Differently from Noselli et al. (2010), a constant value for the material fringe constant $f_\sigma$ has been considered here since
non constant values were found not to introduce significant improvements.
}
 is given by (Frocht, 1965)
\beq
\label{N_true}
N = \frac{t}{f_\sigma}\Delta \sigma,
\eeq
where
$t$ is the sample thickness,
$\Delta \sigma=\sigma_I- \sigma_{II}$ is the in-plane principal stress difference, and
$f_\sigma$ is the material fringe constant, measured by us to be equal to 0.203 N/mm
(using the so-called \lq Tardy compensation procedure', see Dally and Riley, 1965).
These comparisons are reported in Figs. \ref{rettangolazzi} and \ref{rombazzi},
where the full-field solution obtained in Section \ref{campopieno} has been used under plane stress assumption and $\nu = 0.49$.
This assumption is consistent with the reduced thickness of the employed samples, much thinner than the thickness of the samples
employed by Noselli et al. (2010), who have compared photoelastic experiments considering plane strain.
\begin{figure}[!htcb]
  \begin{center}
\includegraphics[width=10 cm]{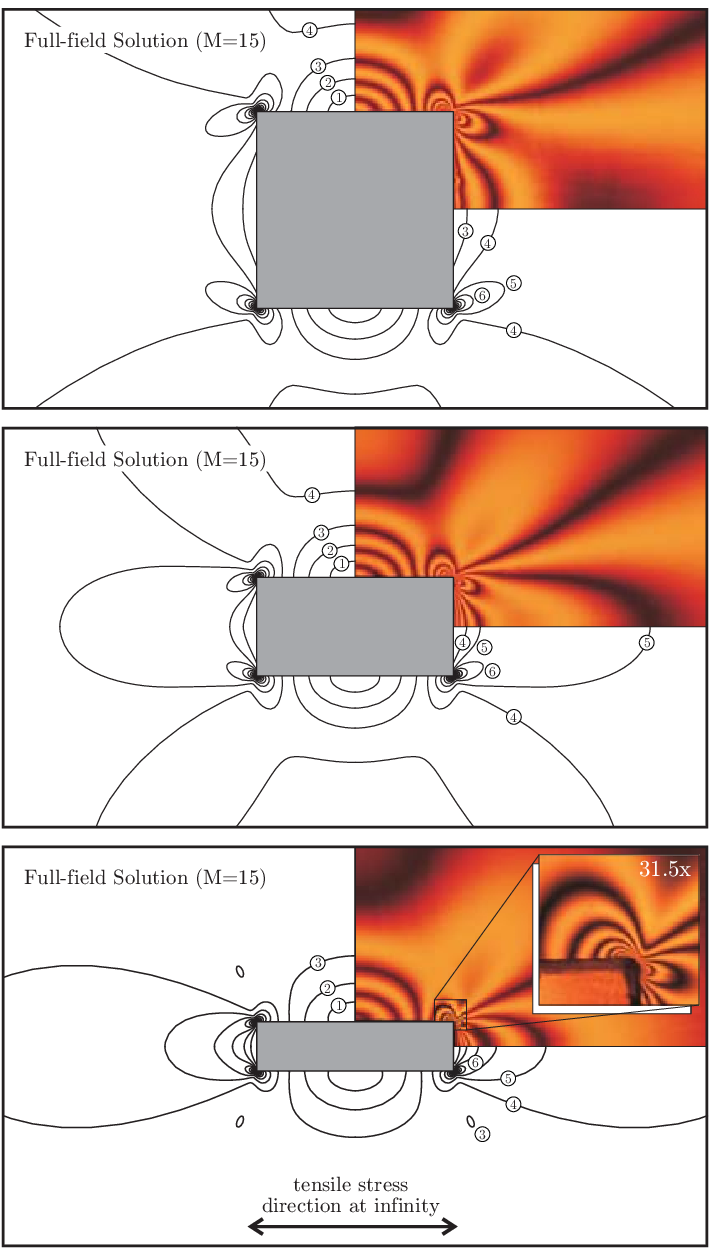}
\caption{
Monochromatic photoelastic fringes (with order number enclosed in a circle) revealing the in-plane principal stress difference field
near stiff rectangular inclusions (made up of polycarbonate, with large edge $l_x=$20 mm and aspect ratios $l_y/l_x=\left\{1;1/2;1/4\right\}$)
embedded in an elastic
matrix (a two-component \lq soft' epoxy resin, approximatively 100 times less stiff than the
inclusion) compared to the elastic solution for rigid inclusions (in plane stress, with Poisson's ratio equal to 0.49), at remote  uniaxial stress
$\sigma_{xx}^\infty=$0.28 MPa (0.22 MPa for the lower part).
}
\label{rettangolazzi}
 \end{center}
\end{figure}
\begin{figure}[!htcb]
  \begin{center}
\includegraphics[width=10 cm]{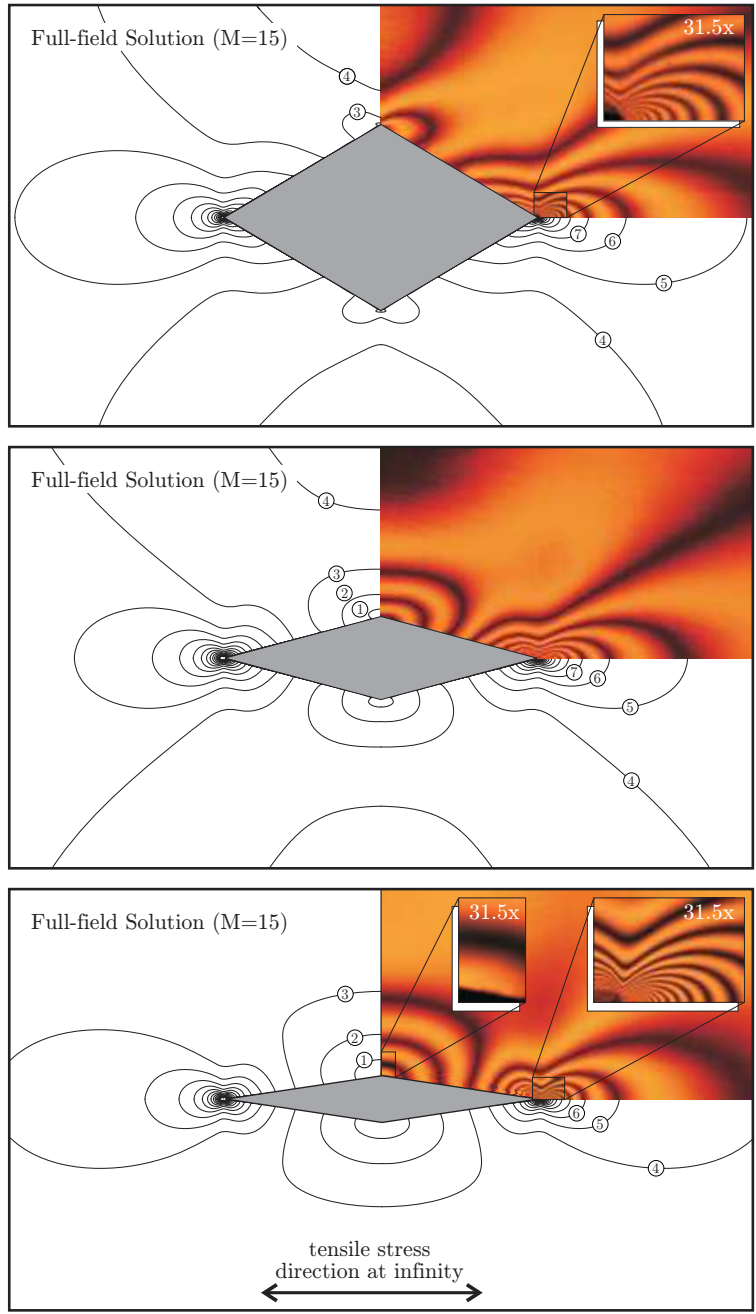}
\caption{
Monochromatic photoelastic fringes (with order number enclosed in a circle) revealing the in-plane principal stress difference field near
stiff rhombohedral inclusions (made up of polycarbonate, with large axes 30 mm and axis aspect ratios $l_y/l_x=\left\{9/15;4/15;2/15\right\}$)
embedded in an elastic
matrix (a two-component \lq soft' epoxy resin, approximatively 100 times less stiff than the inclusion)
compared to the elastic solution for rigid inclusions (in plane stress, with Poisson's ratio equal to 0.49), at remote uniaxial stress
$\sigma_{xx}^\infty=$0.28 MPa (0.25 MPa for the lower part).
} \label{rombazzi}
 \end{center}
\end{figure}

The results show an excellent agreement between theoretical predictions and photoelastic measures,
with some discrepancies near the contact with the inclusions where, the plane stress assumption
becomes questionable due to the out-of-plane constraint imposed by the contact with the rigid phase.\footnote{
The out-of-plane kinematical restriction changes the curvature of the external surface of the sample and has therefore
implications on the use of the method of caustics
for the analysis of rigid inclusions. Moreover, the method of the caustics does not provide a detailed description of the stress fields
around the inclusion, which is nowadays possible with photoelasticity.
Note that the only two contributions to photoelasticity for rigid inclusions published prior to the present article
are those by Gdoutos (1982)
and Theocaris and Paipetis (1976b, their Fig. 9d). The former work is limited only to the analytical description of the fields that
photoelastic experiments
would display, while only one photo of a very poor quality (relative to a  rigid line inclusion) is reported in the latter one.
}
Moreover, microscopical views (at 31.5$\times$) near the vertices of the inclusions,
shown in the inselts of Figs. \ref{rettangolazzi} and \ref{rombazzi}, reveal
that the stress fields are in good agreement even close to the corners, where a strong stress magnification is evidenced near acute corners, while no singularity is observed near obtuse corners.

The near-corner stress magnifications and comparisons with the full field solution (evaluated with $M$ =15) are provided in Fig. \ref{singolarita2},
where the in-plane stress difference (divided by the far field stress) is plotted along the
major axis of the thin and thick rhombohedral inclusions (Fig. \ref{singolarita2}, upper
 and central parts, respectively) and
along a line tangent to the corner (and inclined at an angle $\pi/6$) of the rectangular thin inclusion. In particular,
magnification factors of 5.3 (upper part, $l_y/l_x=2/15$ and $\alpha \approx 23 \pi/24$),
3.8 (central part, $l_y/l_x=9/15$ and $\alpha \approx 5 \pi/6$), and 2.7 (lower part, $l_y/l_x=1/4$ and $\alpha =3 \pi/4$) have been measured.

It is interesting to note that according to the theoretical prediction (Section \ref{asintotototo}, Fig. \ref{singolarita}), the singularity is
stronger for acute than for obtuse inclusion's angles and that the stress fields tend to those corresponding
to a zero-thickness rigid inclusion (a \lq stiffener', see Noselli et al. 2010),
when the rectangular (Fig. \ref{rettangolazzi}) and the rhombohedral
(Fig. \ref{rombazzi}) inclusions become narrow (from the upper part to the lower part of the figures).

According to results shown in Fig. \ref{singolarita}, we observe from Figs. \ref{rettangolazzi}, \ref{rombazzi}, and \ref{singolarita2} the following.

\begin{itemize}

\item  For Mode I loading the stress concentration becomes weak for angles $\alpha$
within $[\pi/2, 3\pi/4]$, see Fig. \ref{rombazzi} (compare  the fields near the two different vertices).

\item For Mode II loading the stress concentration is much stronger than for Mode I.
Stress concentrations generated for mixed-mode at an angle $\alpha = 3\pi/4$ are visible in Fig. \ref{rettangolazzi} near the corners of rectangular
 inclusions.
These concentrations are visibly stronger than those near the wider corner in Fig. \ref{rombazzi} (upper part), which is subject to Mode I;

\item The stress fields evidence boundary layers close to the inhomogeneity, see lower part of
Figs. \ref{rettangolazzi} and \ref{rombazzi}: These boundary layers are crucial in defining
detachment mechanisms and failure modes. Therefore, the shape of a
thin inclusion has an evident impact in limiting the working stress of a mechanical piece in which it is embedded. This conclusion has implications
in the design of material with thin and stiff reinforcements, which can be enhanced through an optimization of the inclusion shape.

\end{itemize}

\begin{figure}[!htcb]
  \begin{center}
\includegraphics[width=6 cm]{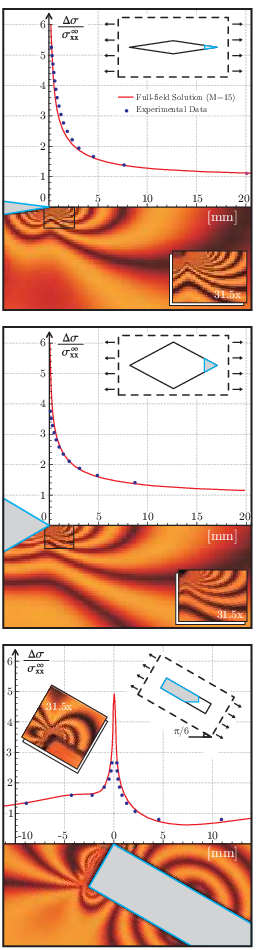}
\caption{
Near-corner stress magnification (in-plane stress difference divided by the far field stress)
for rhombohedral (upper and central parts, respectively $\{l_y/l_x=2/15;\alpha \approx 23 \pi/24\}$ and $\{l_y/l_x=9/15; \alpha \approx 5 \pi/6\}$) and
rectangular (lower part, $l_y/l_x=1/4$ and $\alpha =3 \pi/4$) rigid inclusions.  Experimental results are compared with
the full-field elastic solution, evaluated with $M=15$.
Magnification factors of 5.3 (upper part), 3.8 (central part), and 2.7 (lower part) are visible.
}
\label{singolarita2}
 \end{center}
\end{figure}

\section{Conclusions}

Photoelastic experimental investigations have been presented showing that the stress
field near a stiff inclusion embedded in a soft matrix material can  effectively be
calculated by employing the model of rigid inclusion embedded in a linear elastic
isotropic solid. The results provide also the experimental evidence
of boundary layers, depending on the inhomogeneity shape, which affect the stress fields and therefore define
detachment mechanisms and failure modes.
Finally, the presented methodology paves the way to the experimental stress analysis of
more complex situations, for instance involving interaction between cracks or pores and inclusions as induced by mechanical and thermal loading.

\vspace*{5mm}
\noindent
{\sl Acknowledgments }
DM and DB acknowledge support from EU grant PIAP-GA-2011-286110. FDC and SS acknowledge support from EU grant PIAPP-GA-2013-609758.

{\footnotesize
\noindent
}



\end{document}